\documentclass[epj]{svjour}

\usepackage{graphicx}
\begin{document}
\title{Static Electric Dipole Polarizabilities of Na Clusters}
\author{S. K\"ummel\inst{1}
\and T. Berkus\inst{2} 
\and P.-G. Reinhard\inst{2}
\and M.\ Brack\inst{1}
}                   

\institute{Institute for Theoretical Physics, University of Regensburg,
D-93040 Regensburg, Germany \and 
Institute for Theoretical Physics, University of Erlangen,
D-91077 Erlangen, Germany}
\date{Received: date / Revised version: date}

\abstract{The static electric dipole polarizability of $\mathrm{Na_N}$ clusters
  with even N has been calculated in a collective, axially averaged
  and a three-dimensional, finite-field approach for $2\le N \le 20$,
  including the ionic structure of the clusters. The validity of a
  collective model for the static response of small systems is
  demonstrated. Our density functional calculations verify the trends
  and fine structure seen in a recent experiment. A pseudopotential
  that reproduces the experimental bulk bond length and atomic energy
  levels leads to a substantial increase in the calculated
  polarizabilities, in better agreement with experiment. We relate
  remaining differences in the magnitude of the theoretical and
  experimental polarizabilities to the finite temperature present in
  the experiments.
} 
\maketitle
\section{Introduction}
\label{intro}

 The measurement
of the static electric polarizability of sodium clusters \cite{knight}
and its interpretation in terms of the jellium model \cite{ekardt} was
one of the triggers for the research activities that today form the
field of modern metal cluster physics. The first theoretical studies
were followed by several others with different methods and aims:
density functional calculations using pseudopotentials
\cite{moullet1,moullet2} or taking all electrons into account
\cite{guan} aimed at a quantitative description of the experimentally
observed effects, semi-classical approaches \cite{brack89} focused on
size-dependent trends, and the static electric polarizability served
to test and compare theoretical concepts
\cite{sicforpol,guetxc,chelikowsky}. Recently, the field received new
inspiration 
from a second experimental determination of the static polarizability of
small, uncharged Na clusters \cite{rayane}.

Whereas a qualitative understanding of the experiments can be obtained
with relatively simple models, a quantitative theoretical
determination of the polarizability requires knowledge of the ionic
and electronic configurations of the clusters. Great effort has been
devoted in the past to determine these
\cite{moullet2,martinsalt,koutecky,roethlis}. However, taking all
ionic and electronic degrees of freedom into account in a
three-dimensional calculation is a task of considerable complexity.
Therefore, most of these studies were restricted to clusters with not
more than nine atoms. To reduce the computational expense,
approximations for including ionic effects were developed
\cite{saps,spiegelmann,ppstoer,manninenhueckel,bmcaps,newpp}. A second
problem, however, is the great number of close-lying 
isomers that are found in sodium clusters. This effect is especially
pronounced when stabilization of an overall shape
through electronic shell effects is weak, i.e. for the ``soft'' clusters
that are found between filled shells. In the present work we present
calculations for the static electric polarizability that include the
ionic structure in a realistic way. We take into
account a great number of isomers for clusters with up to 20 atoms,
especially for the soft clusters that fill the second electronic shell.
The theoretical
concepts that we used in this study are introduced in section
\ref{theo}, where we also discuss the relevant cluster structures. In
section \ref{results} we present our results and compare with other
calculations and experimental work. Our conclusions are summarized in
section \ref{conc}.

\section{Theoretical concepts}
\label{theo}

The starting point for the theoretical determination of the
polarizability of a cluster is the calculation of the ionic and
electronic configuration of the ground-state and close lying
isomers. In the present work, this was done in two steps. First, we
calculated low-energy ionic geometries for a wide 
range of cluster sizes with an improved version \cite{newpp} of the
``Cylindrically Averaged Pseudopotential Scheme'' (CAPS)
\cite{bmcaps}. In CAPS the ions are treated fully three-dimensionally,
but the valence electrons are restricted to axial symmetry.
The cluster ground state is found by simultaneously minimizing the
energy functional with respect to the set of ionic positions
(simulated annealing) and the valence-electron 
density. For the exchange and correlation energy we used the
local-density approximation (LDA) functional of Perdew and Wang
\cite{pw}, and for the pseudopotential we employed the recently
developed phenomenological smooth-core potential that reproduces low
temperature bulk and atomic properties \cite{newpp}. Detailed
comparisons with {\it ab initio} calculations have 
shown \cite{newpp} that CAPS predicts ionic geometries
of sodium clusters rather accurately since truly triaxial deformations
are rare. Furthermore, in a second step we performed
fully three-dimensional (3D) Kohn-Sham (KS) calculations to check the
ordering of isomers and to calculate polarizabilities without axial
restriction on the electrons, and also included configurations 
from 3D geometry optimizations into our analysis as discussed below.

\begin{figure}[b]
  \includegraphics[width=\columnwidth]{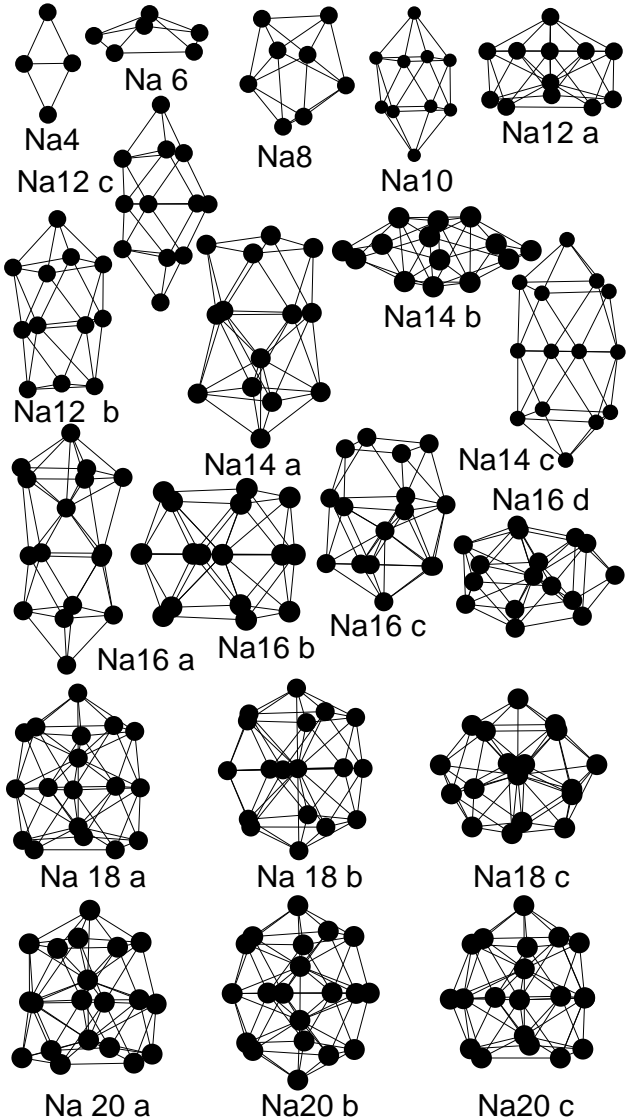}
\caption{Cluster structures $\mathrm{Na_4}$ to
  $\mathrm{Na_{20}}$. See text for discussion.}
\label{structs} 
\end{figure}
Fig. \ref{structs} schematically depicts the most important ionic
geometries for neutral clusters with even electron numbers between 2
and 20. (We have calculated the polarizabilities also for many further
and higher isomers which, however, are not shown in Fig.\
\ref{structs} for the sake of clarity. They were omitted from the
discussion  since they do not lead to qualitatively different
results.) For the small clusters $\mathrm{Na_2}$, $\mathrm{Na_4}$ and
$\mathrm{Na_6}$, many other theoretical predictions have been made
\cite{moullet1,moullet2,rayane,koutecky,roethlis}, and our
geometries are in perfect agreement with them.  In
addition, due to the construction of the pseudopotential \cite{newpp},
the bond lengths are close to the experimental ones, as e.g. seen in
the dimer, where our calculated bond length is $5.78 a_0$ and the
experimental one \cite{expdimer} is $5.82 a_0$. For $\mathrm{Na_8}$
and $\mathrm{Na_{10}}$, our results are in agreement with
3D density functional calculations
\cite{moullet1,moullet2,roethlis}. For $\mathrm{Na_{12}}$, we do not
now of any {\it ab initio} calculations. Therefore, besides two low-energy
configurations from 
CAPS [(a) and (b)], we also included a locally re-optimized low-energy
geometry from a 3D, H\"uckel model calculation
\cite{spiegelmann} in our analysis (c). Our 3D calculations
confirm our CAPS results and find structures (a) and (b) quasi
degenerate with a difference in total energy of 0.05 eV, whereas structure
(c) is higher by 0.4 eV. Two of the three geometries
considered for $\mathrm{Na_{14}}$ [(b) and (c)] were also found very
similar in 3D H\"uckel model calculations
\cite{spiegelmann,manninenhueckel}, and both CAPS and the 3D KS calculations
find all of them very close in energy. For
$\mathrm{Na_{16}}$, we find as the CAPS-ground state structure
(a), and in our 3D calculations structure (b) is quasi degenerate with
(a), whereas structures (c) and (d) are higher by 0.08 eV and 0.5
eV. Due to their very different 
overall shapes, these isomers span a range of what can be expected for
the polarizability. For $\mathrm{Na_{18}}$ and $\mathrm{Na_{20}}$, our
structures are again in close agreement with the 3D
density functional calculation of \cite{roethlis}, and all three structures
are quasi degenerate.

The static electric polarizability was calculated in two different
ways. The first is based on a collective description of electronic
excitations. It uses the well known equality
\begin{equation}
  \alpha=2 m_{-1},
\end{equation}
which relates the negative first moment
\begin{equation}
\label{extmoment}
m_{-1}({\bf Q})=\int_0^\infty E^{-1} S_{\bf Q}(E) \, dE = \sum_\nu
(\hbar \omega_\nu)^{-1} \left| \langle \nu |{\bf Q} | 0 \rangle
\right|^2
\end{equation} 
of the strength function
\begin{equation}
  S_{\bf Q}(E)=\sum_\nu \left| \langle \nu |{\bf Q}| 0 \rangle
  \right|^2 \delta(E_\nu - E_0 -E),
\end{equation}
to the static electric polarizability $\alpha$ in the direction
specified by the external (dipole) excitation operator ${\bf Q}$.

In the evaluation of the strength function, the excited states $|\nu
\rangle$ are identified with collective excitations. A 
discussion of this approach can be found in \cite{lrpa}. The
collective calculations were carried out using the cylindrically
averaged densities and the ``clamped nuclei
approximation'' \cite{moullet2}, {\it i.e.} the ionic positions were
taken to be the same with and without the dipole field. We have also
checked this widely used approximation in the context of our studies
and find it well justified, as discussed below.

The static polarizability can also be calculated directly from the
derivative of the induced dipole moment \boldmath $ \mu$ \unboldmath
in the presence of an external electric dipole field $\bf F$ (``finite
field method''):
\begin{equation}
  \alpha_{ij}=\frac{\mu_j(+F_i)-\mu_j(-F_i)}{2 F_i}, \hspace{0.5cm}
  i,j=x,y,z,
\end{equation}
where
\begin{equation}
  \mu_j({\bf F})=-e \int r_j n({\bf r, F})\,d^3r + e Z \sum_{\bf R}
  R_j
\end{equation}
for ions with valence $Z$.  Here one has to make sure that the
numerically applied finite dipole field $\bf F$ is small enough to be
in the regime of linear response, but that it is on the other hand
large enough to give a numerically stable signal.  We have carefully
checked this and found that the used field strengths between $0.00001
e/{a_0}^2$ and $0.0005 e/{a_0}^2$ meet both requirements.
Applied to the axial
calculations, this approach allows to obtain the polarizability in the
$z$-direction. By employing this method with the 3D KS calculations we have
checked the influences of the axial averaging and the 
collective model on the polarizability and found that the
z-polarizabilities from the axial and the
3D finite-field calculations agree within 1\% on the
average for the low-energy isomers. This shows that the axial
averaging is a good 
approximation for the clusters discussed here. The performance of the
collective model will be discussed below in Section \ref{comptheo}.
The orientation of our coordinate system was
chosen such that the z-axis is in that principal direction of
the tensor of inertia in which it deviates most from its average
value. The average static electric polarizability
\begin{equation}
  \bar{\alpha}:=\frac{1}{3} \, tr(\alpha)
\end{equation}
of course is independent of the choice of coordinate system.

\section{Results}
\label{results}

\subsection{Comparison of different theoretical results}
\label{comptheo}

Since all density functional calculations that we know of agree on the
geometry of the smallest sodium clusters, these clusters can serve as
test cases to compare different theoretical approaches. In Table
\ref{compothers} we have listed the averaged 
static dipole polarizability as obtained in different calculations,
together with the value obtained in the recent experiment of Rayane
{\it et al.} \cite{rayane}.
All calculations reproduce the experimental trend and give the correct
overall magnitude. But also, all calculations underestimate the
polarizability. The magnitude of this underestimation, however, varies
considerably for the different approaches. Whereas our results
are closest to the experiment and close to the theoretical ones of
Ref.\ \cite{rayane}, with the largest difference to the experiment
being 8 \% for $\mathrm{Na_{8}}$, a difference of 27 \% is found for
this cluster in the calculation based on the {\it ab initio} Bachelet,
Hamann, Schl\"uter (BHS) Pseudopotential \cite{moullet2}. A good part
of this difference can be explained by comparing the bond lenghts of
the clusters. The BHS pseudopotential considerably underestimates the
bond lengths \cite{moullet2}, leading to a higher electron density and
a lower polarizability. Our empirical smooth-core pseudopotential, 
on the other hand, was constructed to reproduce the
experimental low-temperature bulk bond length (together with the
compressibility and the atomic $3s$-level) when used with the LDA, and
correspondingly results in a higher polarizability, in better
agreement with experiment. It is further interesting to note that also
the polarizabilities calculated with the empirical Bardsley potential
\cite{moullet1}, which was constructed to reproduce atomic energy
levels, are noticeably higher than the BHS-based values.  This shows
that the cluster polarizability is also sensitive to atomic energy
levels, and the fact that our values are closest to the experiment
thus is a natural consequence of the combination of correct atomic
energy levels and bond lengths.
\begin{table}[htb]
\caption{Averaged static electric polarizability of small sodium
  clusters in ${\mathrm \AA}^3$. Brd: density
  functional (DF) calculation with empirical non-local Bardsley
  pseudopotential \cite{moullet1}. BHS: DF calculation with {\it ab initio}
  non-local Bachelet, Hamann, Schl\"uter
  pseudopotential \cite{moullet2}. All el.: all-electron DF calculation
  including gradient corrections to the exchange-correlation
  functional \cite{guan}.  TM: DF calculation with Troullier-Martins
  non-local pseudopotential \cite{chelikowsky}. GAUSS.: DF calculation
  based on the GAUSSIAN94 program with SU basis set
  \cite{rayane}. Present: present work, values from 
  three-dimensional approach. ExpR: recent experiment \cite{rayane}.}
\label{compothers}      
\begin{tabular}{l|ccccccc}
  \hline\noalign{\smallskip} & Brd. & BHS & All el. & TM & GAUSS. & Present
  & ExpR\\ \noalign{\smallskip}\hline\noalign{\smallskip}
  $\mathrm{Na_{2}}$& 37.7 & 33.1 & 35.9 & 36.2  & 38.2   & 37.0  & 39.3 \\ 
  $\mathrm{Na_{4}}$& 76.3 & 67.1 & 71.4 & 77.2  & 78.4   & 78.7  & 83.8 \\ 
  $\mathrm{Na_{6}}$& 100.3& 89.4 & 94.8 & not   & 104.4  & 107.3 & 111.8 \\ 
  $\mathrm{Na_{8}}$& 111.7& 97.0 & not  & 117.6 & 119.2  & 123.0 & 133.6 \\ 
  \noalign{\smallskip}\hline
\end{tabular}
\end{table}

From comparison with the calculations
that went beyond the LDA \cite{guan,rayane}, it however becomes clear
that the empirical pseudopotentials by construction "compensate" some
of the errors that are a consequence of the use of the LDA. Therefore,
it would be dangerous to argue that the inclusion of gradient
corrections, which have been shown to increase the polarizability,
could bring our calculated values in agreement with experiment: going
beyond the LDA but keeping the empirical LDA pseudopotentials could lead
to a double counting of effects. We therefore conclude that, on the
one hand, a considerable part of the earlier observed differences
between theoretical and experimental polarizabilities can be
attributed to effects associated with errors in the bond lengths or
atomic energy levels, but on the other hand, further effects must
contribute to the underestimation with respect to experiment. We will
come back to this second point below.

In Table \ref{compours} we have listed the polarizabilities of
$\mathrm{Na_{2}}$ to $\mathrm{Na_{20}}$ for the geometries shown in
Fig. \ref{structs}. The left half gives the polarizability as computed
from the 3D electron density with the finite-field
method, and the right half lists the values obtained in the axially
averaged collective approach. For the clusters up to
$\mathrm{Na_{8}}$, the two methods agree well and the differences for
the averaged polarizabilities are less than 1\% for $\mathrm{Na_{2}}$
and $\mathrm{Na_{8}}$, and 3\% for $\mathrm{Na_{4}}$ and
$\mathrm{Na_{6}}$. This shows that the collective description is
rather accurate, which is remarkable if one recalls that we are
dealing with only very few electrons. Beyond $\mathrm{Na_{8}}$, the
differences are 6 \% on the average, 
which is still fair, but obviously higher. This looks
counter-intuitive at first sight, because the collective description
should become better for larger systems. However, for $N>8$
there comes an increasing number of  particle-hole states close to the
Mie plasmon resonance \cite{lrpa}, leading to increasing fragmentation of
the collective strength. $m_{-1}$ and
thus $\alpha$ is sensitive to energetically low-lying excitations
since their energies enter in the denominator in Eq.\ \ref{extmoment},
and this can lead to an underestimation of the polarizability.
\begin{table}[htb]
\caption{Static electric polarizability in ${\mathrm \AA}^3$ for the cluster
  geometries of Fig. \ref{structs}. Left half: three-dimensional,
  finite field calculation. Right half: cylindrically averaged,
  collective calculation.}
\label{compours}      
\begin{tabular}{l|rrrr|rrr}
\hline\noalign{\smallskip} & \multicolumn{4}{c}{3D, finite field}&
\multicolumn{3}{c}{cyl., coll. mod.} \\ 
& $\alpha_x$ & $\alpha_y$ & $\alpha_z$ & $\bar{\alpha}$ &
$\alpha_\rho$ & $\alpha_z$ &$ \bar{\alpha}$ \\ 
\noalign{\smallskip}\hline\noalign{\smallskip}
$\mathrm{Na_{2}}$ & 29.9 & 29.9 & 51.3 & 37.0 & 30.2 & 51.5 & 37.3
\\ 
$\mathrm{Na_{4}}$ & 47.0 & 59.0 & 130.1 & 78.7 & 53.7 & 122.6 & 76.7
\\ 
$\mathrm{Na_{6}}$ & 129.8 & 129.8 & 62.2 & 107.3 & 124.9 & 62.1 & 103.9 
\\ 
$\mathrm{Na_{8}}$ & 118.1 & 118.5 & 132.3 & 123.0 & 117.5 & 131.4 & 122.1
\\ 
$\mathrm{Na_{10}}$ & 126.3 & 126.3 & 219.3 & 157.3 & 125.3 & 194.5 & 148.4 
\\ 
$\mathrm{Na_{12}a}$ & 213.4 & 213.4 & 155.7 & 194.2 & 199.5 & 143.6 & 180.9
\\ 
$\mathrm{Na_{12}b}$ & 156.5 & 158.6 & 261.1 & 192.1 & 151.3 & 234.7 & 179.1
\\ 
$\mathrm{Na_{12}c}$ & 158.6 & 145.1 & 285.3 & 196.3 & 148.7 & 252.0 & 183.1 
\\ 
$\mathrm{Na_{14}a}$ & 183.5 & 183.5 & 278.9 & 215.3 & 177.5 & 251.7 & 202.2 
\\ 
$\mathrm{Na_{14}b}$ & 271.3 & 273.9 & 137.8 & 227.7 & 264.2 & 132.7 & 220.4 
\\ 
$\mathrm{Na_{14}c}$ & 175.2 & 179.3 & 291.3 & 215.3 & 171.1 & 262.5 & 201.6 
\\ 
$\mathrm{Na_{16}a}$ & 193.5 & 193.5 & 394.5 & 260.5 & 190.3 & 318.4 & 233.0 
\\ 
$\mathrm{Na_{16}b}$ & 235.2 & 235.2 & 231.5 & 234.0 & 227.3 & 220.3 & 225.0 
\\ 
$\mathrm{Na_{16}c}$ & 212.5 & 213.4 & 272.2 & 232.7 & 206.0 & 255.1 & 222.4 
\\ 
$\mathrm{Na_{16}d}$ & 272.2 & 260.8 & 239.3 & 239.3 & 260.6 & 185.1 & 235.4
\\ 
$\mathrm{Na_{18}a}$ & 260.3 & 262.5 & 291.3 & 271.4 & 236.4 & 266.7 & 246.5 
\\ 
$\mathrm{Na_{18}b}$ & 251.8 & 250.7 & 283.8 & 262.1 & 232.6 & 253.4 & 239.5 
\\ 
$\mathrm{Na_{18}c}$ & 281.6 & 280.5 & 250.7 & 270.9 & 255.9 & 227.7 & 246.5 
\\ 
$\mathrm{Na_{20}a}$ & 285.7 & 284.5 & 309.4 & 293.2 & 269.7 & 279.7 & 273.0 
\\ 
$\mathrm{Na_{20}b}$ & 275.0 & 275.0 & 311.8 & 287.3 & 261.8 & 282.9 & 268.8 
\\ 
$\mathrm{Na_{20}c}$ & 267.9 & 271.5 & 295.2 & 278.2 & 283.2 & 280.3 & 282.2 
\\ \noalign{\smallskip}\hline
\end{tabular}
\end{table}

Comparing the polarizabilities of clusters with the same number of
electrons but different geometries shows the influence of the overall
shape of the cluster. For $\mathrm{Na_{14}}$, e.g., isomers (a) and (c) have a
valence electron density which is close to prolate, whereas (b) has a
more oblate one. The averaged polarizability for the two prolate
isomers is equal, although their ionic geometries differ. The oblate
isomer, however, has a noticeably higher averaged polarizability. This
is what one expects, because for oblate clusters there are two
principal directions with a low and one with a high polarizability,
whereas for prolate clusters the reverse is true. The fact that
different ionic geometries can lead to very similar averaged
polarizabilities is also seen for $\mathrm{Na_{12}}$. It thus becomes
clear that contrary to what was believed earlier \cite{moullet1} one  
cannot necessarily distinguish between details in the ionic
configuration by comparing theoretical values to experimental data
that measure the averaged polarizability.

\subsection{Comparison with experiments}
\label{compexp}

\begin{figure}[bth]
  \includegraphics[angle=270,width=\columnwidth]{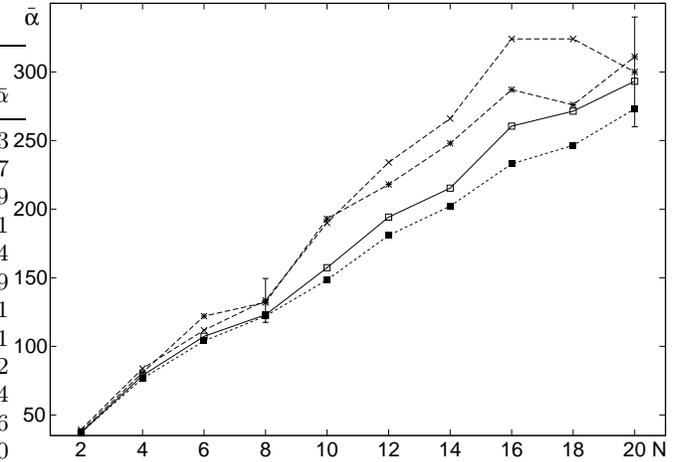}
\caption{Static electric dipole polarizability in $\mathrm \AA^3$
  versus number of electrons. 
  Crosses with thin, long dashed line: experiment of Rayane {\it et al.}
  \cite{rayane}; 
  stars with strong, long dashed line: experiment of Knight {\it et
    al.} \cite{knight}; 
  open squares with full line: present work, three-dimensional finite-field
  calculation for lowest isomer; 
  filled squares with short dashed line: present work, axially averaged
  collective calculation for lowest isomer. See text for discussion of
  error bars.} 
\label{abspol} 
\end{figure}

\begin{figure}[bt]
  \includegraphics[angle=270,width=\columnwidth]{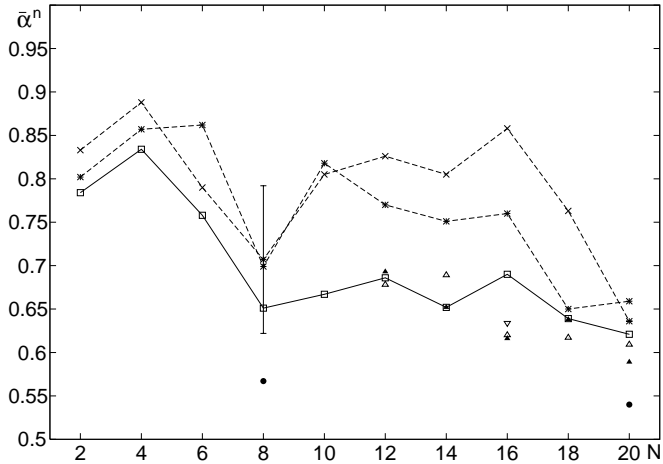}
\caption{Normalized static electric dipole polarizability. 
  Crosses with thin dashed line: experiment of Rayane {\it et
    al.} \cite{rayane}; stars with strong dashed line: experiment of Knight
  {\it et al.} \cite{knight}; 
  squares with full line: present work, three-dimensional 
  finite-field calculation for lowest isomer; open triangle, filled
  triangle and upside-down triangle: second, third and fourth isomer,
  respectively; filled circles: Jellium results from
  \cite{ekardt,guetxc}. See text for discussion.   
  }
\label{normpol} 
\end{figure}

Fig.\ \ref{abspol} shows $\bar{\alpha}$ for our ground state
structures as obtained in the axial, collective approach and the
3D finite-field calculations, in comparison to the two
available sets of experimental data. The absolute values for the
experiments were calculated from the measured relative values with an
atomic polarizability of $23.6 \mathrm \AA^3$ \cite{rayane}. To guide the eye,
the polarizabilities from each set of data are connected by lines.
Both experiments and the theoretical data show that, overall, the
polarizability increases with increasing cluster size. The
polarizability from the axial collective model qualitatively shows the
same behavior as the one from the 3D finite field calculation.
Comparison of the 3D values with the experimental data shows
that for the smallest clusters, the theoretical and experimental
values agree as discussed before, and the values obtained in the two
experiments are comparable up to $\mathrm{Na_{10}}$. Beyond
$\mathrm{Na_{10}}$, the discrepancies between the two experiments
become larger, and also the differences between theoretical and
experimental polarizabilities increase. For $\mathrm{Na_{12}}$,
$\mathrm{Na_{14}}$, $\mathrm{Na_{16}}$ and $\mathrm{Na_{18}}$ the
experiment of Knight {\it et al.} gives lower values than the
experiment of Rayane {\it et al.}, and the calculated averaged
polarizability is lower than both experiments for $\mathrm{Na_{12}}$,
$\mathrm{Na_{14}}$, and $\mathrm{Na_{16}}$. For $\mathrm{Na_{18}}$ the
finite-field value obtained for our ground-state structure
matches the value measured by Knight {\it el al.}, and for
$\mathrm{Na_{20}}$, our ground-state polarizability is very close to
the measurement of Rayane {\it et al.} In this discussion one must
keep in mind, however, that the experimental uncertainty is about +/-2 $\mathrm
\AA^3$ per atom \cite{rayane}, i.e. the uncertainty in the absolute
value increases with the cluster size, as indicated by the errors bars
in Fig.\ \ref{abspol}.  
Comparisons are made easier if the linear growth in $\bar{\alpha}$ is
scaled away. Therefore, one should rather look at the normalized polarizability
\begin{equation}
\label{defnormpol}
\bar{\alpha}^{\rm n} := \frac{\bar{\alpha}}{N \alpha_{\mathrm{atom}}},
\end{equation}
which is shown in Fig.\ \ref{normpol}, because it allows to identify
trends and details more clearly.

From Fig. \ref{normpol} it becomes clear that for $\mathrm{Na_{2}}$ to
$\mathrm{Na_{8}}$, the trend seen in the two experiments is similar up
to one exception: For $\mathrm{Na_{6}}$, the experiment of Rayane {\it
  et al.} predicts a noticeably smaller value than the one by Knight
{\it et al.} Comparison with our theoretical data shows that, although
the values of the older experiment are closer to the theory with
respect to magnitude for $\mathrm{Na_{2}}$, $\mathrm{Na_{4}}$ and
$\mathrm{Na_{8}}$, the trend that our data show corresponds clearly to
the one seen in the new experiment since the two curves are parallel.
Going from $\mathrm{Na_{8}}$ to $\mathrm{Na_{10}}$, both experiments
predict a steep rise in the polarizability. This rise due to the shell
closing at $\mathrm{Na_{8}}$ is also seen in the theoretical data, but
it is less pronounced than in the experiments (as we will discuss
below). For $\mathrm{Na_{12}}$, a higher $\bar{\alpha}^{\rm n}$ than for
$\mathrm{Na_{10}}$ is predicted by the data of Rayane, whereas the
reverse ordering is seen in the data of Knight {\it et al.} Again, our
calculations support the finding of the new experiment, and all
isomers lead to similar $\bar{\alpha}^{\rm n}$. For $\mathrm{Na_{14}}$, both
experiments show a decrease. Our prolate ground state and isomer
reproduce this trend. That it is the prolate structures that fit to
the experiment is consistent with the 
{\it ab initio} molecular dynamics calculations of H\"akkinen {\it et
  al.} \cite{hakkinen}. The next step to $\mathrm{Na_{16}}$ again
reveals a slight difference between the two experiments: both predict
an increase compared to $\mathrm{Na_{14}}$, but whereas the older
experiment sees $\bar{\alpha}^{\rm n}$ smaller for $\mathrm{Na_{16}}$ than
for $\mathrm{Na_{12}}$, the new experiment shows the opposite
ordering. Once more, our ground state structure leads to a
polarizability that follows the trend of the new experiment. (The other
isomers, however, lead to smaller polarizabilities, and an explanation
for the difference between the experiments thus might be that
different ensembles of isomers were populated due to slightly
different experimental conditions.) Going to
$\mathrm{Na_{18}}$ leads 
to a decrease in the polarizability in both experiments. Our
calculation shows this decrease, which is a manifestation of the
nearby shell closing. But whereas the old experiment actually sees the
shell closing at $\mathrm{Na_{18}}$ and an increase in the
polarizability for $\mathrm{Na_{20}}$, the new experiment and our data
find an absolute minimum at $\mathrm{Na_{20}}$.

A comparison with the polarizability obtained in the spherical jellium
model \cite{ekardt,guetxc} for $\mathrm{Na_{8}}$ and
$\mathrm{Na_{20}}$, also indicated in Fig.\  
\ref{normpol}, shows the improvement that is brought about by the
inclusion of the ionic structure.

\subsection{Discussion}

As just discussed, our calculations reproduce the fine structure seen
in the new experiment. However, there is no obvious explanation for
why the results of the two experiments differ \cite{private}. Also,
there is a characteristic change in the magnitude of the difference
between our theoretical results and the experimental values of Rayane
{\it et al.}: whereas the calculated values for $\mathrm{Na_{2}}$ to
$\mathrm{Na_{8}}$ and $\mathrm{Na_{20}}$ on the average differ only by
5 \% from the new experiment, the open-shell clusters from
$\mathrm{Na_{10}}$ to $\mathrm{Na_{18}}$ show 18 \% difference for the
ground state. The increase in polarizability when going from
$\mathrm{Na_{8}}$ to $\mathrm{Na_{10}}$ is considerably
underestimated, whereas the following steps in the normalized
polarizability are nearly reproduced correctly, i.e. it looks as if
the theoretical curve for $\mathrm{Na_{10}}$ to $\mathrm{Na_{18}}$
should be shifted upwards by a constant. A first suspicion might be
that this ``offset'' could be due to the use of CAPS in the geometry
optimization. But it should be noted that the step occurs at
$\mathrm{Na_{10}}$, and that also $\mathrm{Na_{14}}$ and
$\mathrm{Na_{18}}$ are off by the same amount. Since the low-energy
structures of these clusters are well established, as discussed in
Section \ref{theo}, and since also the geometry for $\mathrm{Na_{12}}$
from the 3D calculation does not lead to
qualitatively different results, we can conclude that the differences
are not due to limits in the geometry optimization with CAPS. Also,
the neglection of the relaxation of the nuclei in the presence of the
electric dipole field has been investigated earlier 
\cite{moullet2} for the small clusters and was shown to be a good
approximation. We have counter checked this result for the test case Na$_{10}$
and find corrections of less than 1\%.

An obvious limitation of our approach is the neglect of the core
polarization. However, the all-electron calculations of Guan {\it et
  al.} \cite{guan} treat the core electrons explicitly and do not lead to
better agreement with experiment, as discussed in
Section \ref{comptheo}. From this, one already can conclude that core
polarization cannot account for all of the observed differences. Its
effect can be estimated from the polarizability of the sodium cation.
Different measurements \cite{corepol} find values between
$0.179\,\mathrm \AA^3$ and $0.41 \,\mathrm \AA^3$, leading to
corrections of - roughly - 1-2\,\% in $\bar{\alpha}^{\rm n}$. Since the core
polarizability leads to a 
shift in $\bar{\alpha}^{\rm n}$ that is the same for all cluster sizes, it
contributes to the difference that is also seen for the smallest
clusters, but it cannot explain the jump in the difference seen at
$\mathrm{Na_{10}}$.

Another principal limitation of our approach is the use of the LDA.
As, e.g., discussed in \cite{guan}, the LDA can affect the
polarizability in different and opposing ways. On the one hand it may
lead to an overscreening and thus an underestimation of the
polarizability, and early calculations within the spherical jellium
model reported that indeed the static polarizability was increased if
one went beyond LDA using self-interaction corrections
\cite{sicforpol}. On the other hand, self-interaction corrections can
lead to more negative single particle energies and thus to smaller
polarizabilities \cite{guan}, and Refs.\ \cite{guetxc} and
\cite{ullrich} give examples where the overall effect of
self-interaction corrections on the optic response is very small. One
cannot directly conclude from the jellium results to our ionic
structure calculations, because the sharp edge of the steep-wall
jellium model can qualitatively lead to differences. But in any case
it is highly implausible that the LDA affects the clusters from
$\mathrm{Na_{10}}$ to $\mathrm{Na_{18}}$ much stronger than the other
ones, and it should also be kept in mind that the worst indirect
effects of the LDA as, e.g., underestimation of bond lengths, are
compensated by using our phenomenological pseudopotential.

One might also ponder about possible uncertainties in the
experimental determination of the polarizabilities.
A considerable underestimation could be explained if one assumes that
while passing through the deflecting field, the clusters are oriented
such that one always measures the highest component of the
polarizability. In that case, we would not have to compare the
averaged value to the experiment, but the highest one. One could
imagine that the cluster's rotation be damped, since angular momentum
conservation is broken by the external field and the energy thus could
be transferred from the rotation to internal degrees of freedom
(vibrations). The time scale of this energy transfer is not known, but
since the clusters are spending about $10^{-4}$s in the deflecting
field region, it seems unlikely that there should be no coupling over
such a long time. One could further argue that for statistical reasons
it is less likely that a larger cluster will lose its orientation
again through random-like thermal motion of its constituent ions than
a smaller cluster. However, the maximal energy difference between
different orientations is very small, for $\mathrm{Na_{10}}$, e.g., it
is $\mathrm{0.3 K  \, k_B} $ for the typical field strength applied in
the experiment \cite{knight}. Thus, thermal fluctuations can be
expected to wipe out any orientation.

From another point of view, however, the finite temperature explains a
good part of the differences that our calculations (and other
calculations for the small clusters) show in comparison to the
experimental data. Whereas our calculations were done for T=0, the
supersonic nozzle expansion used in the experiment produces clusters
with an internal energy distribution corresponding to about 400 - 600
K \cite{hansen,durgourd}. An estimate based on the thermal expansion
coefficient of bulk sodium leads to an increase in the
bond lengths of about 3 \%, and a detailed finite-temperature CAPS
calculation \cite{baerbel} for $\mathrm{Na_{11}^+}$ at 400 K also
shows a bond length increase of 3 \%. This will only be a lower limit,
since in neutral clusters one can expect a larger expansion than in
the bulk due to the large surface, and also a larger expansion than
for charged clusters. Thus, to get an estimate for the lower limit of
what can be expected from thermal expansion, we have scaled the
cluster coordinates by 
3 \% and again calculated the polarizabilities, finding an increase of
about 3 \% for the planar and 5 \% for  
three-dimensional structures. This finding is consistent with the
results of Guan {\it et al.} Together with the corrections that are to be
expected from the core polarizability, this brings our results for the
small and the closed shell clusters in quantitative agreement with the
experimental data.

\section{Summary and Conclusion}
\label{conc}

We have presented calculations for the static electric dipole
polarizability for sodium clusters with atom numbers between 2 and 20,
covering several low-energy structures for each cluster size beyond
$\mathrm{Na_{10}}$. By comparing our results to previous calculations
for the smallest clusters, we have shown that a pseudopotential which
correctly reproduces atomic and bulk properties also improves the
static response considerably. We have shown that a collective model
for the excited states of sodium clusters, whose validity for the
dynamical response was established previously, works reasonably also
for the static response in realistic systems. Over the whole range of
cluster sizes studied in the present work, we confirm the fine
structure seen in a recent experiment. By comparing the calculated
averaged polarizability of different isomers for the same cluster size
to the measured polarizability, we showed that completely different
ionic geometries can lead to very similar averaged
polarizabilities. By considering higher isomers we furthermore took a
first step to take into account the finite temperature present in the
experiment. Our results show that for the open shell clusters from
$\mathrm{Na_{10}}$ to $\mathrm{Na_{18}}$, also higher lying isomers do
not close the remaining gap between theory and experiment. This shows
that it is a worthwhile task for future studies to investigate the
influence of finite temperatures on these ``soft'' clusters
explicitly. For $\mathrm{Na_{2}}$ to $\mathrm{Na_{8}}$ and
$\mathrm{Na_{20}}$, we showed that quantitative agreement is already
obtained when the effects of thermal expansion and the core
polarizability are taken into account.

\vspace{1cm}

\small{
One of us (S. K\"ummel) thanks K.\ 
Hansen for several clarifying discussions concerning the experimental
temperatures and time-scales, especially with respect to the
``orientation question'', and the Deutsche Forschungsgemeinschaft for
financial support.
}

\end{document}